\documentstyle[prl,aps,epsfig]{revtex}

\def\eeq{\end{equation}}
\def\beq{\begin{equation}}
\def\bea{\begin{eqnarray}}
\def\eea{\end{eqnarray}}
\begin{document}
\title{Asymmetric unimodal maps at the edge of chaos}
\author{Ugur Tirnakli$^{1}$, Constantino Tsallis$^{2}$ and
Marcelo L. Lyra$^{3}$}
\address{$^1$Department of Physics, Faculty of Science, Ege University, 35100
Izmir, Turkey \\
$^2$Centro Brasileiro de Pesquisas Fisicas, Rua Xavier Sigaud\\
150, 22290-180 Rio de Janeiro-RJ, Brazil\\
$^3$Departamento de F\'{\i}sica, Universidade
Federal de Alagoas, 57072-970 Macei\'o - AL, Brazil\\
tirnakli@sci.ege.edu.tr, tsallis@cbpf.br, marcelo@fis.ufal.br}
\maketitle

\begin{abstract}
We numerically investigate the sensitivity to initial conditions of 
asymmetric unimodal maps $x_{t+1} = 1-a|x_t|^{z_i}$ ($i=1,2$ correspond 
to $x_t>0$ and $x_t<0$ respectively, $z_i >1$, $0<a\leq 2$, $t=0,1,2,...$) 
at the edge of chaos. We employ three distinct algorithms to characterize 
the power-law sensitivity to initial conditions at the edge of chaos, 
namely: direct measure of the divergence of initially nearby trajectories,
the computation of the rate of increase of generalized nonextensive entropies 
$S_q$ and multifractal analysis. The first two methods provide consistent 
estimates for the exponent governing the power-law sensitivity. 
In addition to this, we verify that the multifractal analysis does not provide 
precise estimates of the singularity spectrum $f(\alpha)$, specially near 
its extremal points. Such feature prevents to perform a fine check of the 
accuracy of the scaling relation between $f(\alpha)$ and the entropic index 
$q$, thus restricting the applicability of the multifractal analysis for 
studing the sensitivity to initial conditions in this class of asymmetric maps.
\\

\noindent
{\it PACS Number(s): 05.45.-a, 05.20.-y, 05.70.Ce}
\end{abstract}



\vspace{1.5cm}

\section{Introduction}
In recent years, there has been an increasing interest on the
behaviour of the one-dimensional dissipative maps at their chaos
threshold\cite{TPZ,costa,lyra,circle,latbar2,ugur,ugur2,lyrabjp,singlesite}.
When the sensitivity to the initial conditions is examined at the
onset of chaos, the sensitivity function, defined through

\beq
\xi(t) = \lim_{\Delta x(0)\rightarrow 0}
\frac{\Delta x(t)}{\Delta x(0)}\;\; ,
\eeq
(where $\Delta x(0)$ and $\Delta x(t)$ are the discrepancies of the
initial conditions at times $0$ and $t$), can be put in a conveniently
generalized form

\beq
\xi(t)= \left[1+(1-q)\lambda_q t\right]^{1/(1-q)}\;\;\;\;\;\;
(q\in {\cal R}),
\eeq
(solution of $\dot \xi=\lambda_q \;\xi^q$) where $\lambda_q$ is the 
generalized Lyapunov exponent. This equation
recovers the standard exponential form $\exp{(\lambda_1 t)}$ for $q=1$
(here, $\lambda_1$ is the standard Lyapunov exponent),
but generically $q\neq 1$ corresponds to a power-law behaviour.
In this case, if $\lambda_q < 0$ and $q>1$ ($\lambda_q > 0$ and $q<1$)
the system is said to be {\it weakly} insensitive (sensitive) to the
initial conditions, a situation which is different from the standard case
where we have {\it strong} insensitivity (sensitivity) for $\lambda_1 < 0$
($\lambda_1 > 0$).

Although asymptotic power-law sensitivity to initial conditions was
observed previously\cite{grass,politi,mori}, $\xi(t)$ as given by eq. (2) 
provides a more complete description than just
$\xi(t)\propto t^{1/(1-q)}$ ($t>>1$).
At the edge of chaos, $\xi(t)$ presents strong fluctuations with time,
reflecting the fractal structure of the critical attractor, and eq.(2)
delimits the power-law growth of the upper bounds of the sensitivity
function. These upper bounds allow us to estimate the proper value
$q^*$ of the index $q$ for the map under consideration. This method has
already been successfully used for a variety of one-dimensional dissipative
maps such as logistic\cite{TPZ}, $z$-logistic\cite{costa},
circle\cite{lyra}, $z$-circular\cite{circle} maps.

The second method of estimating the $q^*$ value of the map under
consideration comes from the geometrical aspects of the attractor at
the chaos threshold. This method is based on the multifractal singularity
spectrum $f(\alpha)$, which reflects the fractal dimension of the subset
with singularity strength $\alpha$\cite{multif1,multif2}. The
$f(\alpha)$ function is a down-ward parabola-like concave curve and
typically vanishes at two points, namely $\alpha_{min}$ and $\alpha_{max}$,
characterizing the scaling behavior of the most concentrated and most 
rarefied regions on the attractor.
The study of the scaling behaviour of these regions led two of us to
propose a new scaling relation as\cite{lyra}

\beq
\frac{1}{1-q^*} = \frac{1}{\alpha_{min}} - \frac{1}{\alpha_{max}}\;\; .
\;\;\;\;\;\;\;\;\;  (q^* < 1)
\eeq
This is, in fact, a fascinating relation since it connects the power-law
sensitivity to initial conditions of such dynamical systems with purely
geometrical quantities and consequently it provides a completely different
method for the determination of the proper $q^*$ value of the map under
consideration. This method has also been used so far for logistic\cite{lyra},
$z$-logistic\cite{lyra}, circle\cite{lyra} and $z$-circular\cite{circle} 
maps, and the results obtained for the $q^*$ values are within a good
precision the same as those of the first method.

In order to make the situation even more enlightening, a third method of
obtaining the proper $q^*$ value of a given map has been
introduced very recently using a specific generalization of the
Kolmogorov-Sinai (KS) entropy \cite{latbar2,ugur}. It is known that,
for a chaotic dynamical system, the rate of loss of information can be
characterized by the KS entropy ($K_1$) and it is defined as the increase,
per unit time, of the standard Boltzmann-Gibbs entropy
$S_1=- \sum_{i=1}^{W} p_i \ln p_i$ (we use $k_B=1$). In fact, the KS 
entropy is defined, in principle, through a {\it single-trajectory} in 
phase space based on the frequencies of appearance, in increasingly long 
strips, of symbolic sequences of the regions of the partitioned phase 
space \cite{hilborn}. However, apparently in almost all cases, this 
definition can be equivalently replaced by an ensemble-based procedure, 
which is, no doubt, by far simpler computationally than the former procedure. 
This ensemble-based version is the one we use herein. On the other hand,
it is worth noting that a single-trajectory-based procedure has been
used very recently in \cite{grigo}.

From the Pesin equality, namely, $K_1=\lambda_1$ if $\lambda_1 >0$ and
$K_1=0$ otherwise, it is evident that the KS entropy is deeply related
to the Lyapunov exponents. The KS entropy rate is then defined through
$K_1\equiv \lim_{t\rightarrow\infty}\lim_{W\rightarrow\infty}
\lim_{N\rightarrow\infty} S_1(t)/t$, where $t$ is the time steps, $W$ is
the number of regions in the partition of the phase space and $N$ is the
number of initial conditions (all chosen at $t=0$ within one region among 
the $W$ available ones) that are evolving in time. On the other hand,
for the marginal cases where $\lambda_1=0$, a generalized version of the
KS entropy $K_q$ has been introduced \cite{TPZ} as the increase rate of
a proper nonextensive entropic form, namely

\beq
S_q(t)= \frac{1-\sum_{i=1}^{W} [p_i(t)]^q}{q-1}\; .
\eeq
This entropy enables a generalization of the standard Boltzmann-Gibbs
statistics \cite{ts1,ts2} and it covers the BG entropy as a special case
in the $q\rightarrow 1$ limit. A general review and related subjects on
this nonextensive formalism can be found in \cite{genel}; recent 
applications in high energy physics, turbulence and biology can be seen 
in \cite{bediaga}, \cite{turbulence} and \cite{arpita} respectively. 
Therefore, for the generalized version of KS entropy, the entropy rate 
is proposed to be

\beq
K_q\equiv \lim_{t\rightarrow\infty}\lim_{W\rightarrow\infty}
\lim_{N\rightarrow\infty} \frac{S_q(t)}{t} \; .
\eeq
Consistently, the Pesin equality is also expected to be generalizable as
$K_q=\lambda_q$ if $\lambda_q >0$ and $K_q=0$ otherwise.

Consequently, these ideas have been used very recently to construct a 
third method of estimating the $q^*$ values \cite{latbar2}.
It is conjectured that (i) a unique value of $q^*$ exists such that $K_q$ 
is finite for $q=q^*$, vanishes for $q>q^*$ and diverges for $q<q^*$,
(ii) this value of $q^*$ coincides with that coming from the other two
methods described previously. These conjectures have been verified with
numerical calculations, at the edge of chaos, for the standard logistic
map \cite{latbar2}, logistic-like map family and generalized cosine
map \cite{ugur}, which strongly supports the point that all three
methods yield one and the same special $q^*$ value of a map under
consideration. At this point, it is worth mentioning that when the 
initial conditions are very spread in phase space (instead of the 
localized ones), another class of $q^*$ values (above unity instead 
of below unity) has been found for $z$-logistic case \cite{MTL}.

Although these three different methods of finding $q^*$ value have been
already tested and numerically verified for a number of one-dimensional
dissipative map families, it is no doubt convenient (in the spirit of 
further clarifying their domain of validity)  to test them in more
general grounds. For example, all the maps discussed so far belong to
one-dimensional, dissipative, symmetric, one- or two-parameter unimodal 
families. At this point, one can ask what happens for the
(i) two- (or more) dimensional maps, (ii) conservative maps,
(iii) asymmetric families.
Needless to say, if anyone of these cases could be analyzed with
the above mentioned three methods consistently, the scenario would
obviously become more robust. In the present effort, we shall try
to make a step forward addressing the point (iii), namely the
asymmetric three-parameter family of logistic map of the form

\begin{eqnarray}
 x_{t+1} = f(x_t) \equiv \left\{
\begin{array}{ll}
1- a|x_t|^{z_1} & \mbox{if $x_t \ge 0$} \\
1- a|x_t|^{z_2} & \mbox{if $x_t \le 0$}
\end{array}
\right.
\end{eqnarray}
where $z_{1,2} >1$, $0<a\leq 2$, $-1\le x_t \le 1$ and $t=0,1,2,...~$.

\section{Asymmetric logistic map family: Numerical results}
The properties of this kind of asymmetric map family have already been
studied \cite{jensenma,vieira1,vieira2}. The asymmetric shape of the map
family is illustrated in Fig.~1a for a typical value of ($z_1,z_2$) pair,
whereas in Fig.~1b the bifurcation diagram has been plotted.
Before the onset of chaos, the sequence of bifurcations is the same as
that of Feigenbaum, but in the chaotic region (after the onset of chaos),
the relative sizes of the various windows are quite different from those
of the $z$-logistic map (namely, $z_1=z_2=z$). Moreover, it is
well-known that this map family fails to exhibit the metric universality
of Feigenbaum. In this case, the scaling factors (Feigenbaum numbers)
$\alpha_F$ and $\delta$ present an oscillatory divergent
behaviour \cite{jensenma,vieira1}. Same kind of oscillatory behaviour has
also been observed for multifractal function $f(\alpha )$ \cite{vieira2}.

Since $q^*$ values were not available for asymmetric logistic map
family, it was not possible to see the behaviour of $q^*$ as a function of
the ($z_1,z_2$) pairs. On the other hand, in a very recent
effort \cite{ugur2}, in order to see this behaviour, {\it without finding
the precise values of $q^*$ for ($z_1,z_2$) pairs}, one of us has used
another technique based on the very recent generalization of bit cumulants
for chaotic systems \cite{raman1,raman2}. In spite of the fact that in
$q$-generalized bit cumulant theory, $q$ is a free parameter, it seems
from the results of \cite{ugur2} that as $z_2-z_1\rightarrow \pm \infty$,
$q^*$ will approach unity, which is similar to the behaviour observed for
symmetric maps studied so far \cite{TPZ,costa,lyra,circle}.

We are now prepared to proceed with our numerical results for the
asymmetric logistic map family. First of all, since our aim is to look at
the properties of this family at the onset of chaos, the calculated
values of the critical map parameter ($a_c$) as a function of ($z_1,z_2$)
pairs are given in Fig.~2 and in the Table. It is evident that the behaviour 
of $a_c$ values with respect to ($z_2 - z_1$) is very similar to the tendency 
of $a_c$ values of the $z$-logistic family with respect to parameter $z$.

\subsection{First Method}
As already discussed above, this method is based on the sensitivity to
initial conditions and for the asymmetric logistic map family, the
sensitivity function $\xi(t)$ is given by

\beq
\ln \xi(t) = \sum_{t=1}^l \ln \left[\frac{df(x_t)}{dx}\right]
\eeq
and exhibits, at the chaos threshold, a power-law divergence,
$\xi \propto t^{1/(1-q^*)}$, from where $q^*$ values can be calculated
by measuring, on a log-log plot, the upper bound slope $1/(1-q^*)$.
In Fig.~3, for $x_0=0$, the behaviour of the sensitivity function
has been illustrated for two typical ($z_1,z_2$) pairs. The slope
of the upper bound has been calculated for each pair between the
time interval [4,8.5]. We encountered that, as $(z_2 - z_1)$ values
become larger (that is the map becomes more asymmetric), the number
of points that could be used in estimating the slope becomes fewer.
For such cases, one should go for times larger than say 8.5 
(in logarithmic scale), which requires much more precision on the 
values of $a_c$. On the other hand, for the $(z_2,z_1)$ pairs given 
in the Table, the above mentioned time interval is good enough to 
determine the slope. From this slope, for each pair, we calculate the 
$q^*$ values and in Fig.~4 we exhibit the behaviour of $q^*$ as a function 
of $(z_2 - z_1)$ for two typical pairs. It is seen that as $(z_2 - z_1)$ 
goes $\pm \infty$, $q^*$ becomes closer to unity. This tendency is consistent 
with the recent claim of \cite{ugur2} and also similar to the behaviour of 
symmetric map families studied so far \cite{TPZ,costa,lyra,circle}.

\subsection{Second Method}
As mentioned previously, for this asymmetric family, the multifractal
function $f(\alpha)$ fluctuates considerably for different number of
iterations ($I$) which prevents us to estimate exact values of
$\alpha_{min}$ and $\alpha_{max}$ from where we determine $q^*$ values.
Worse than that, this problem cannot be cured by
extrapolating the number of iterations to infinity as it is done
in \cite{circle} for the $z$-circular maps and \cite{singlesite} for 
the single-site map. 
In $z$-circular case, the fluctuations are so systematic that one
can extrapolate the results to infinite number of iterations with
acceptable precision from where $\alpha_{min}$ and $\alpha_{max}$
values could be deduced, whereas for asymmetric logistic family this
is not the case. To illustrate this, we plotted the $f(\alpha)$ curve
for the inflexion pair $(2,3)$ in Fig~5a, where the oscillatory behaviour
is evident. Moreover, we presented in Fig~5b the extrapolation of
$\alpha_{min}$ and $\alpha_{max}$ for the same pair. It is clear
from the confidence interval that it is not possible to estimate
the correct values of them due to large fluctuations.
This yields us to conclude that for this asymmetric map family, the 
second method cannot be used easily to determine $q^*$ values due to
the unavoidable fluctuations in the $f(\alpha)$ function. In fact, this
result has also been supported by a recent observation: One of us has
shown recently \cite{lyra2} that for $z$-logistic maps, the scaling
relation given in Eq.~(3) can be reexpressed as
$1/(1-q)=[(z-1)\ln2]/[\ln\alpha_F(z)]$, which clearly points out that
this scaling is dependent of Feigenbaum constant $\alpha_F$. Since
for the asymmetric map family we are studying, as already mentioned,
the Feigenbaum numbers exhibit oscillatory divergent
behaviour \cite{jensenma,vieira1}, it is evident that $q^*$ values
cannot be easily and reliably inferred from the scaling relation due to 
these fluctuations.

\subsection{Third Method}
Finally, in order to verify the results of the first method, let us
use the entropy increase rate procedure to estimate the proper
$q^*$ values. The procedure is the following: First, we partition the
phase space into $W$ equal cells, then we choose one of them and
select $N$ initial conditions (all inside the chosen cell). As $t$
evolves, these initial conditions spread within the phase space and
naturally this gives us a set $\{N_i(t)\}$ with $\sum_{i=1}^W N_i(t)=N,
\; \forall t$, which consequently yields a set of probabilities
$\{p_i(t)\equiv N_i(t)/N\}$. In the beginning of time, clearly
$S_q(0)=0$, then it gradually exhibits three successive regions as
firstly indicated in \cite{latbar} for a different system. In the first 
region, the entropy is roughly constant in time, then it starts increasing 
in the second region and finally it tends towards its saturation value. 
This indicates that the linear increase of the proper entropy is expected 
to emerge in the second (intermediate) region.
As clearly explained in \cite{latbar2,ugur}, at the
chaos threshold, very large fluctuations appear in the entropy due
to the fact that the critical attractor occupies only a tiny part of
the available phase space. To overcome this problem, we use a procedure
of averaging over the efficient initial conditions as discussed
in \cite{latbar2,ugur}. Since this procedure is very time-consuming,
we apply it for two typical ($z_1,z_2$) pairs to check the results
of the first method. The results are given in Fig.~6. It is observed
that, for all cases, in the intermediate region, the linear increase
of the entropy with time emerges only for a special value of $q$
(namely $q^*$), and this value corresponds, within a good precision,
to the one obtained from the first method. On the other hand,
for $q<q^*$ ($q>q^*$) it curves upwards (downwards).
To provide quantitative support to this, we fit the curves with the
polynomial $S_q(t)=A+Bt+Ct^2$ in the interval [$t_1,t_2$] characterizing
the intermediate region. The nonlinearity coefficient
$R\equiv C(t_1+t_2)/B$ is a measure of the importance of
the nonlinear term, therefore $R$ vanishes for a strictly linear fit.
These results are given as insets of Fig.~6.

\section{Conclusions}

In this work, we performed an extensive analysis of the sensitivity to
initial conditions problem related to a family of asymmetric maps at the
edge of chaos. We have been particularly interested in exploiting
the connections between the sensitivity function, generalized
non-extensive entropies and the multifractal character of the
critical attractor.

A direct numerical computation of the sensitivity function $\xi(t)$,
which measures the temporal evolution of the distance between initially 
nearby trajectories, shows strong fluctuations whose 
upper bounds delimit a power-law growth $\xi(t)\propto t^{1/(1-q^*)}$. 
The characteristic power-law exponent was determined for
several pairs of the inflexions at the left and right of map inflexion 
point. For extremely asymmetric maps, wild fluctuations do not allow 
the power-law exponents to be determined with high accuracy, but the 
general trend indicates that $q^*$ approaches unity in the limit 
of very asymmetric maps.

We also employed a multifractal analysis, based on the standard 
Halsey {\em et al} algorithm\cite{multif1}, to compute the singularity 
spectrum $f(\alpha )$ related to the critical attractor of the present 
family of asymmetric maps. A recently proposed scaling relation 
associates the extremal points of the  $f(\alpha )$ curve with the 
power-law exponent governing the sensitivity function. However, the 
numerical method used to compute $f(\alpha)$ exhibits large fluctuations 
when applied to these asymmetric maps. This feature makes difficult the 
precise estimate of the location of its extremal points. It would be 
valuable to have an alternative algorithm to compute the $f(\alpha)$ curve 
which could overcome this point to allow a fine check the accuracy of the 
scaling relation (Eq.~3) for these asymmetric maps.

Finally, the sensitivity to initial conditions was investigated by
computing the rate of increase of generalized entropies $S_q$. At
the edge of chaos, there is a particular entropic index $q^*$ for
which the entropy grows, in the infinitely fine-graining limit, at a 
stationary rate after a short initial transient. This method provides 
values for $q^*$ which are in agreement with the ones obtained from the 
direct measure of the sensitivity function. Although being more time 
consuming, the entropy measure is free from wild fluctuations and allows 
for a relatively fine and confident estimate of $q^*$. Therefore, this
method should be the starting point to investigate the possibility
of similar power-law sensitivity to initial conditions in higher
dimensional as well as conservative non-linear dynamical systems 
(see, for instance, \cite{fulvio}).

\section*{Acknowledgments}
One of us (UT) acknowledges the partial support of the Ege
University Research Fund under the project number 2000FEN049.
This work is also partially supported by PRONEX, CNPq, CAPES, FAPEAL 
and FAPERJ (Brazilian agencies).


\newpage


{\bf Table and Figure Captions}

\vspace{1.5cm}

{\bf Table} - The values of $a_c$ and $q^*$ for various ($z_1,z_2$) pairs.\\

{\bf Figure 1} - (a) Asymmetric shape of the map given by Eq.(6) for the
inflexion parameter pair (2,4).
(b) The bifurcation diagram of the map for the same inflexion parameter pair.\\

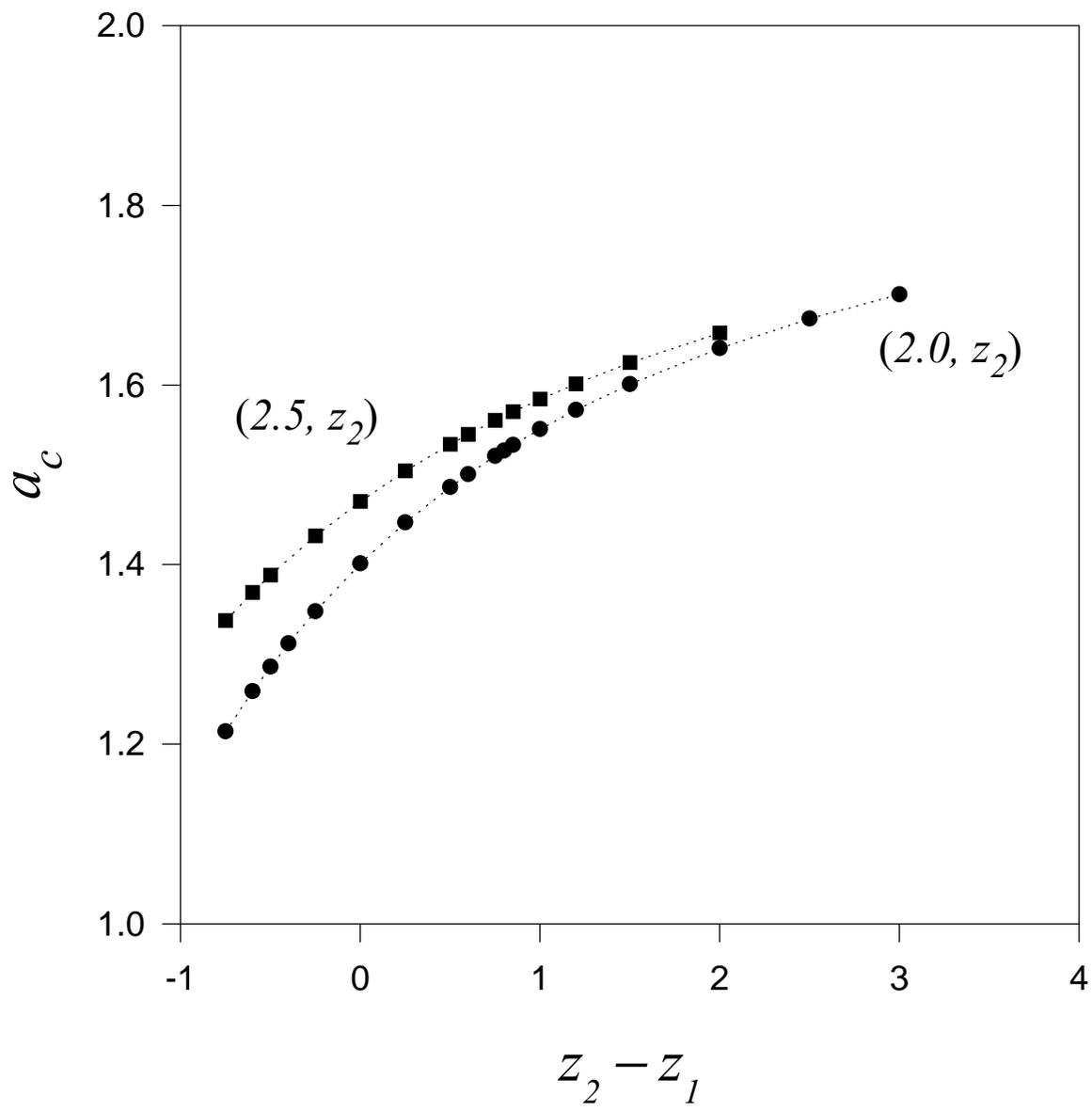
{\bf Figure 2} - The behaviour of the critical map parameter $a_c$ as a
function of ($z_2 - z_1$) for two typical inflexion parameter pairs.
The dotted lines are guides to the eye. \\

{\bf Figure 3} - Log-log plot of the sensitivity function versus time for
(a) (2, 1.75) (b) (2.5, 3) pairs.\\

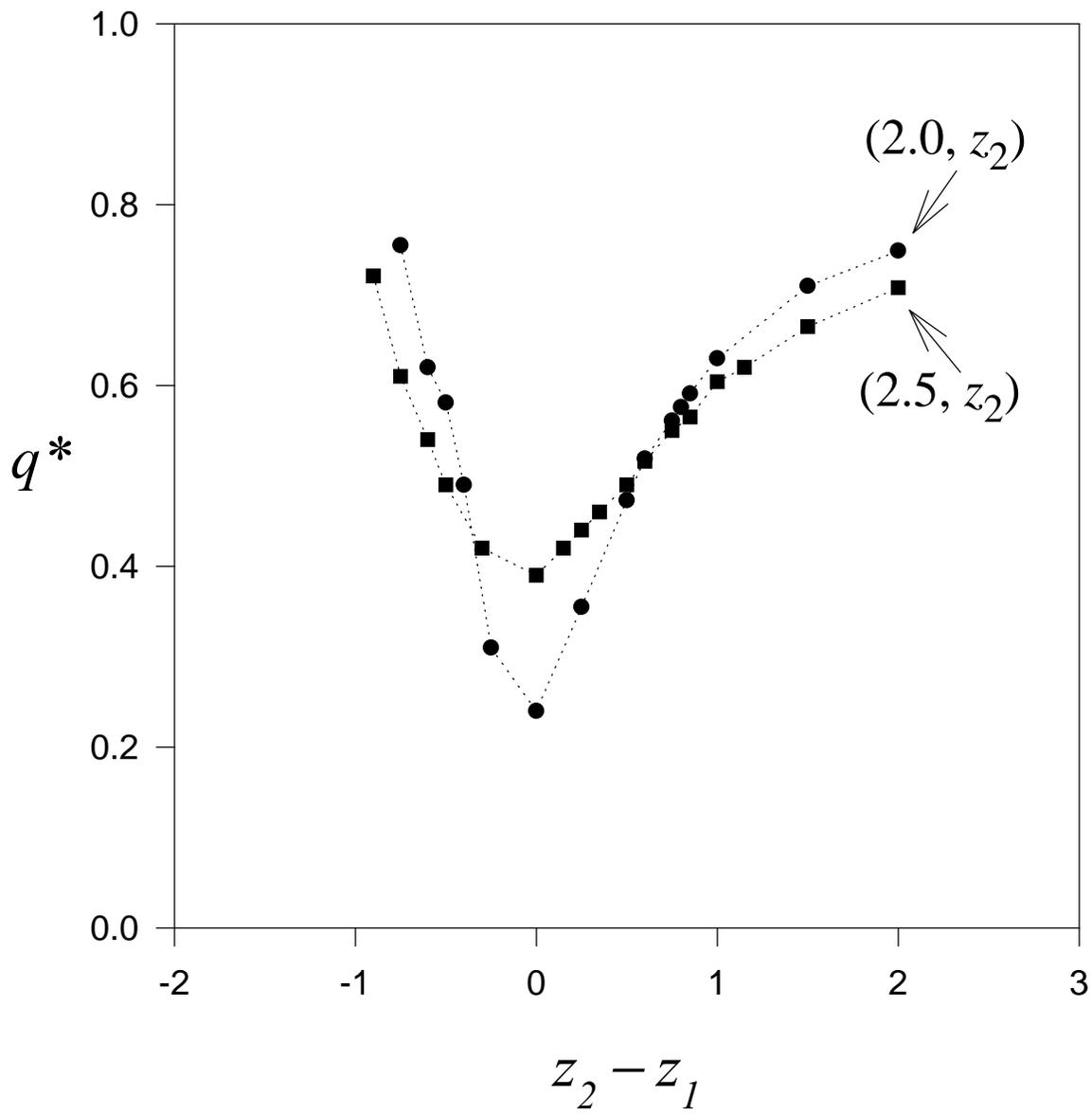
{\bf Figure 4} - The behaviour of $q^*$ values as a function of
($z_2 - z_1$) for ($2, z_2$) and ($2.5, z_2$). The dotted lines are
guides to the eye. See the Table for typical error bars.\\

{\bf Figure 5} - (a) The behaviour of $f(\alpha)$ curve for various
values of the number of iterations. (b) The oscillatory behaviour of
$\alpha_{min}$ and $\alpha_{max}$. The dotted lines are standard confidence 
intervals (SigmaPlot 1.00). \\

{\bf Figure 6} - Time evolution of the entropy for three different values
of $q$ for (a) (2, 1.75) ; (b) (2.5, 3) pairs. Insets: The nonlinearity
coefficient $R$ versus $q$. The interval characterizing the intermediate
region is [13,31] for (a) and [7,25] for (b). The dotted lines are
guides to the eye. \\


\newpage

\begin{center}
{\bf Table}

\vspace{1cm}

\begin{tabular}{||c|c|c||}
\hline
($z_1,z_2$) & $a_c$ & $q^*$ \\ \hline
$\;(2,1.25)\;$ & $\;1.21403412...\;$ & $\;0.76\pm0.01\;$ \\ \hline
$\;(2,1.4)\;$ &  $\;1.25863959...\;$ & $\;0.62\pm0.01\;$ \\ \hline
$\;(2,1.5)\;$ &  $\;1.28613959...\;$ & $\;0.58\pm0.01\;$ \\ \hline
$\;(2,1.6)\;$ &  $\;1.31201155...\;$ & $\;0.49\pm0.01\;$ \\ \hline
$\;(2,1.75)\;$ & $\;1.34799246...\;$ & $\;0.31\pm0.01\;$ \\ \hline
$\;(2,2)\;$  &   $\;1.40115518...\;$ & $\;0.24\pm0.01\;$ \\ \hline
$\;(2,2.25)\;$ & $\;1.44691055...\;$ & $\;0.36\pm0.01\;$ \\ \hline
$\;(2,2.5)\;$ &  $\;1.48645043...\;$ & $\;0.47\pm0.01\;$ \\ \hline
$\;(2,2.75)\;$ & $\;1.52083316...\;$ & $\;0.56\pm0.01\;$ \\ \hline
$\;(2,3)\;$ &    $\;1.55094551...\;$ & $\;0.63\pm0.01\;$ \\ \hline
$\;(2,3.5)\;$ &  $\;1.60109881...\;$ & $\;0.71\pm0.01\;$ \\ \hline
$\;(2.5,1.6)\;$& $\;1.30334301...\;$ & $\;0.72\pm0.01\;$ \\ \hline
$\;(2.5,1.75)\;$&$\;1.33742470...\;$ & $\;0.61\pm0.01\;$ \\ \hline
$\;(2.5,2)\;$ &  $\;1.38805851...\;$ & $\;0.49\pm0.01\;$ \\ \hline
$\;(2.5,2.5)\;$ &$\;1.47055000...\;$ & $\;0.39\pm0.01\;$ \\ \hline
$\;(2.5,3)\;$ &  $\;1.53418776...\;$ & $\;0.49\pm0.01\;$ \\ \hline
$\;(2.5,3.25)\;$&$\;1.56070446...\;$ & $\;0.55\pm0.01\;$ \\ \hline
$\;(2.5,3.5)\;$ &$\;1.58439440...\;$ & $\;0.60\pm0.01\;$ \\ \hline
$\;(2.5,4)\;$ &  $\;1.62488124...\;$ & $\;0.67\pm0.01\;$ \\ \hline
\end{tabular}
\end{center}

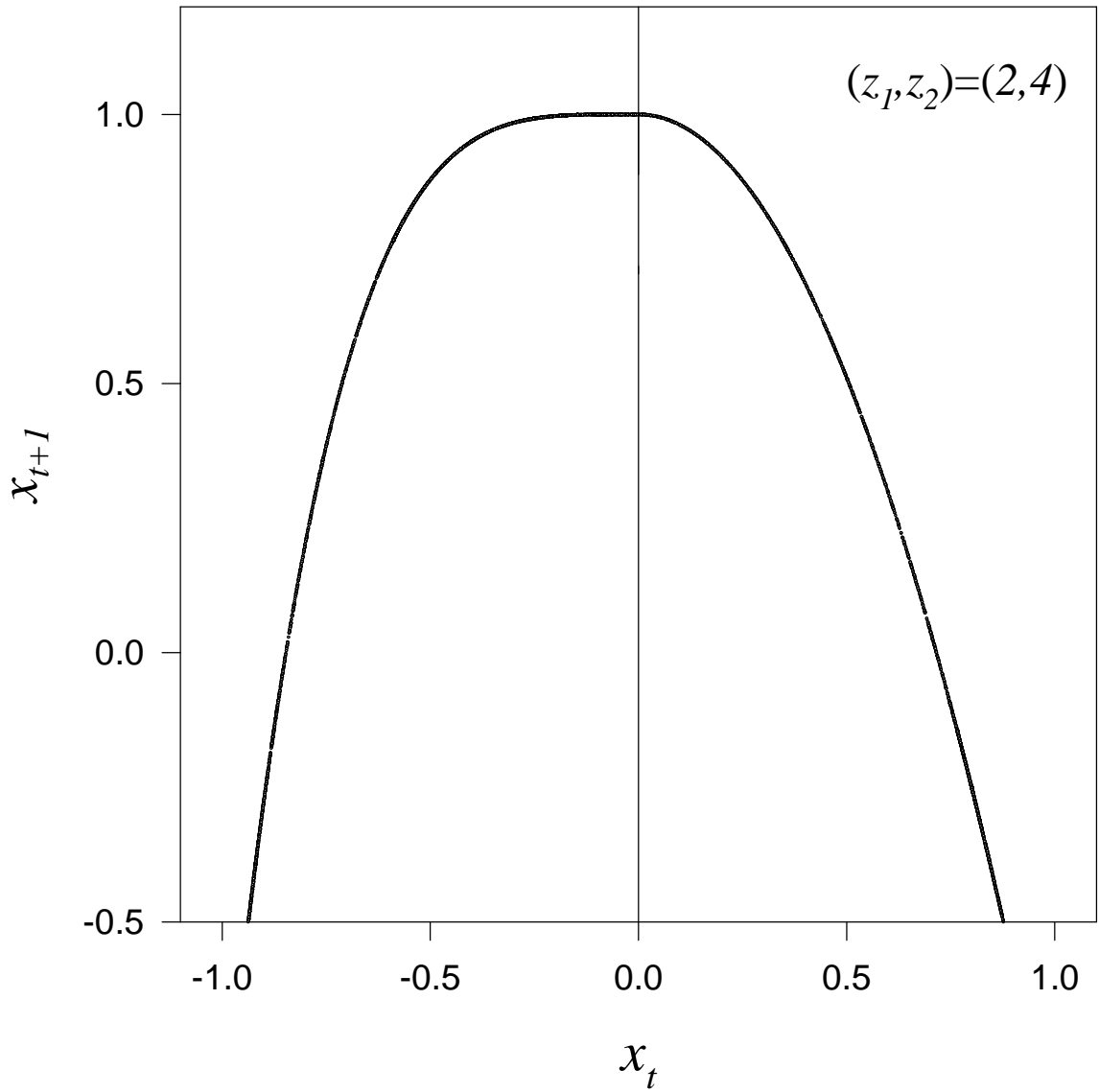
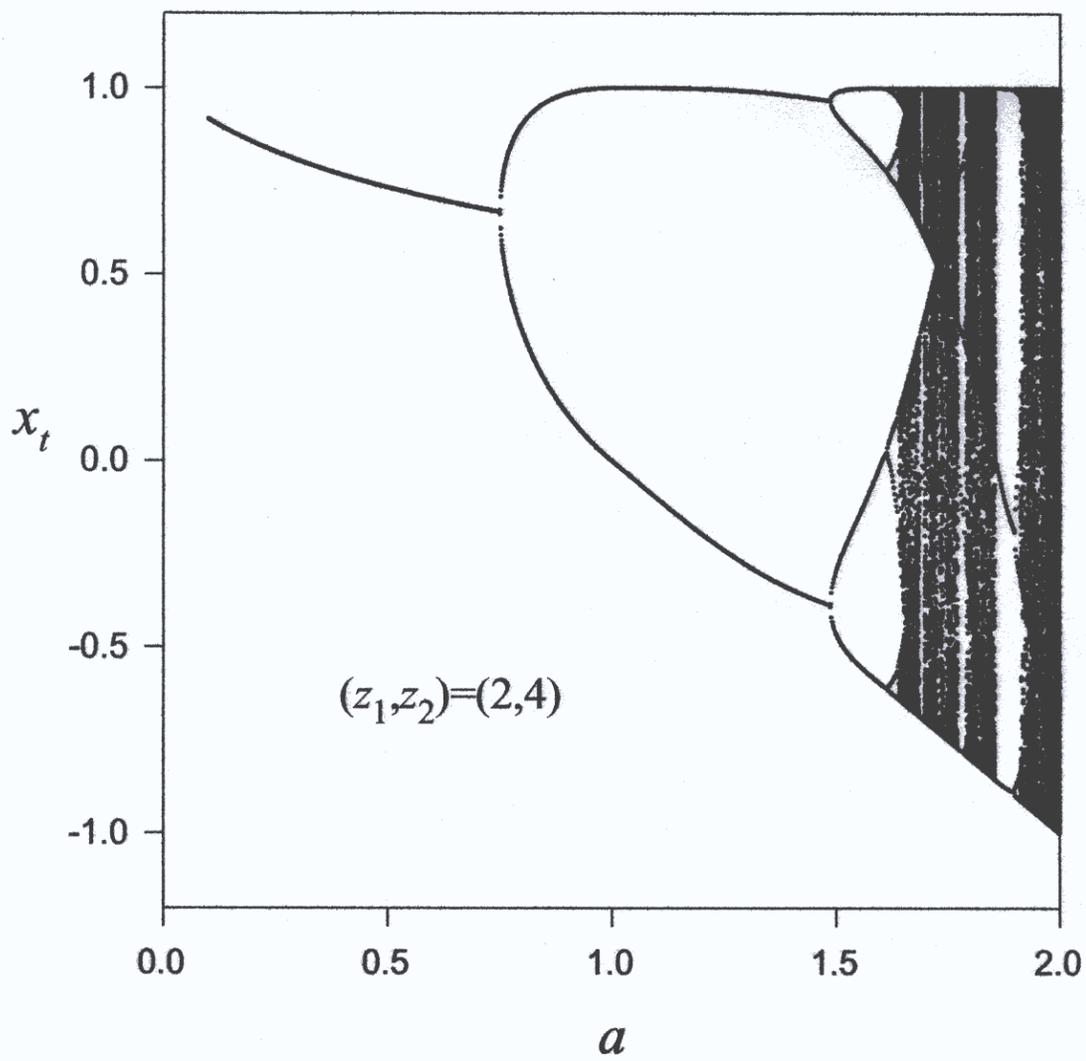
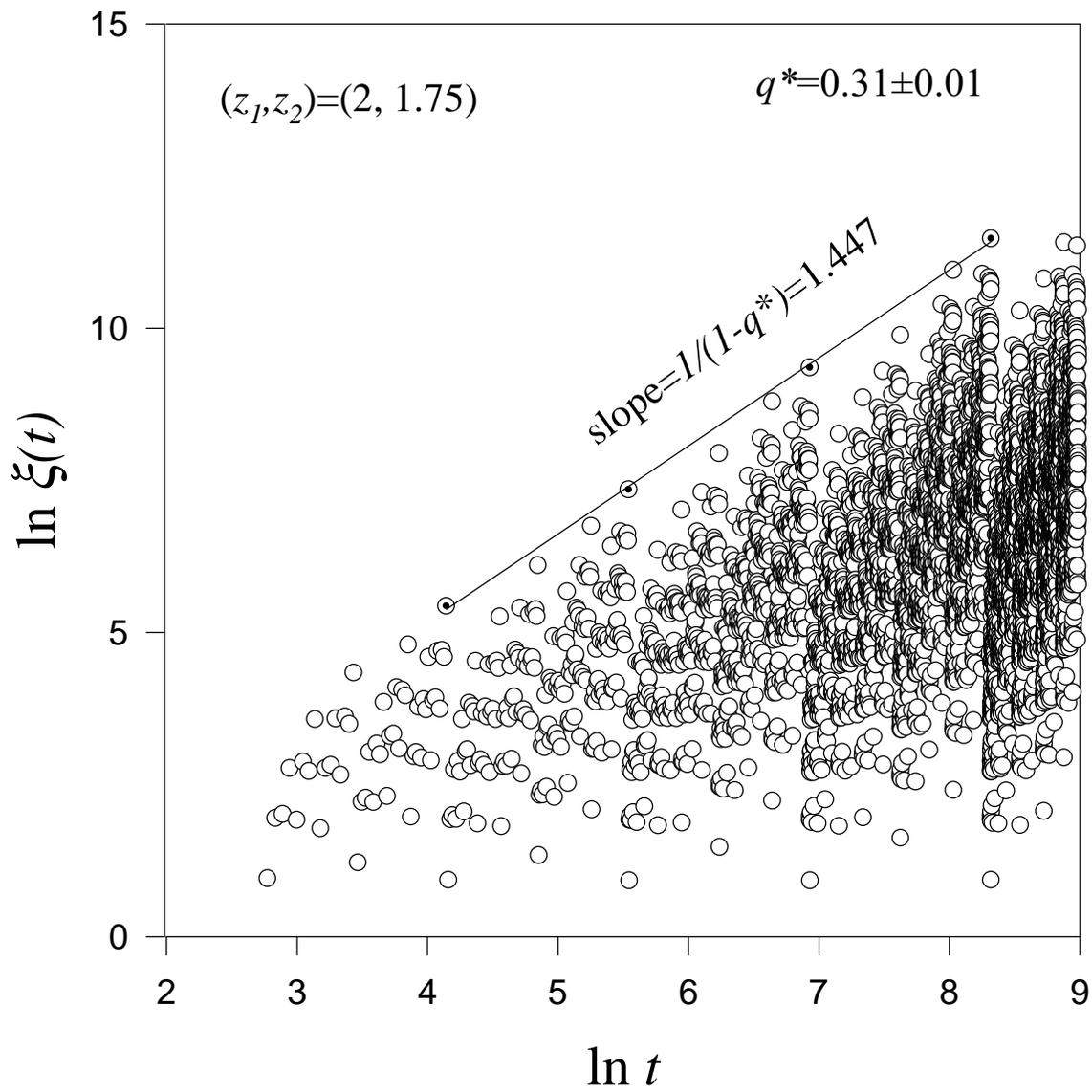
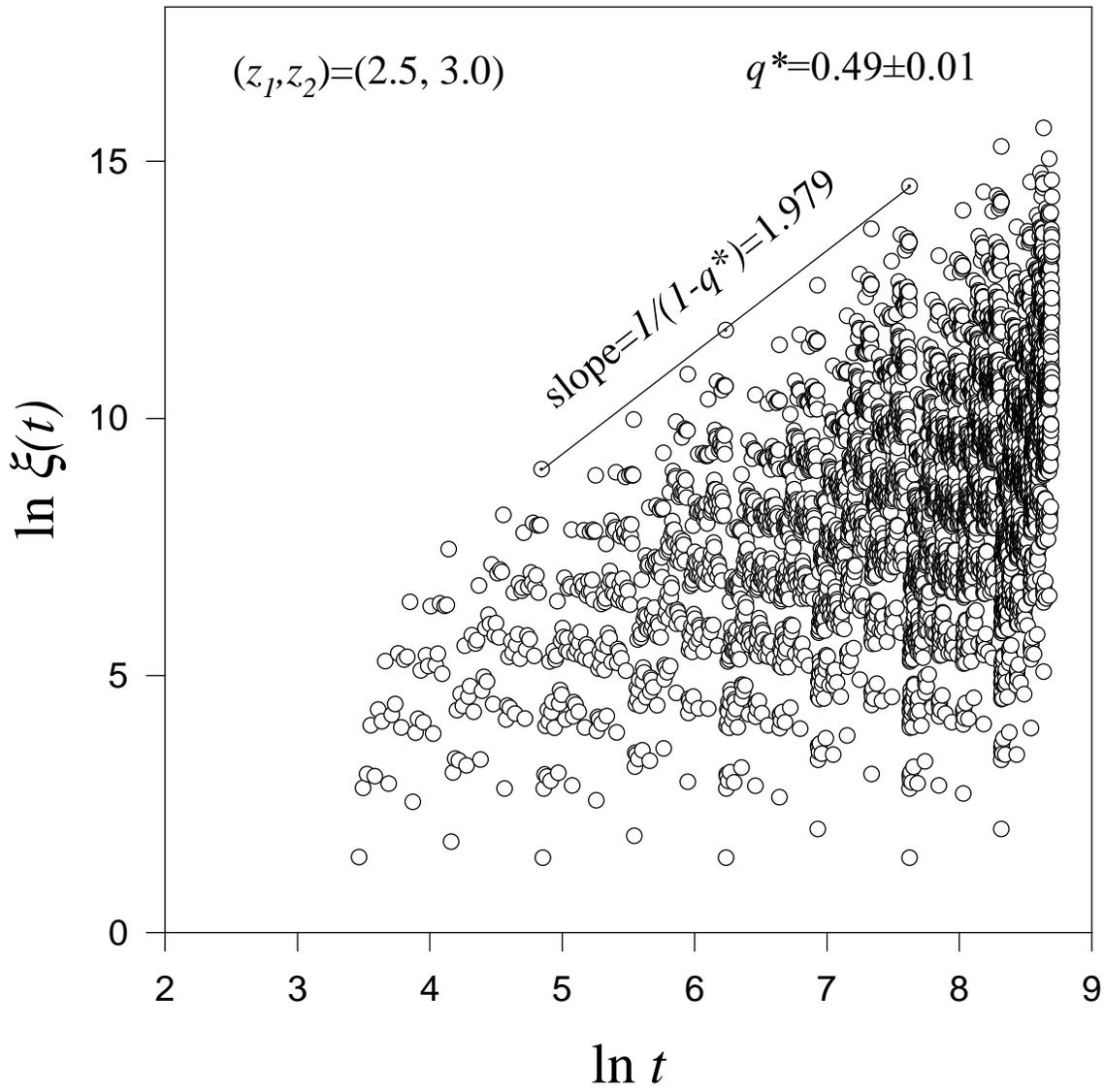
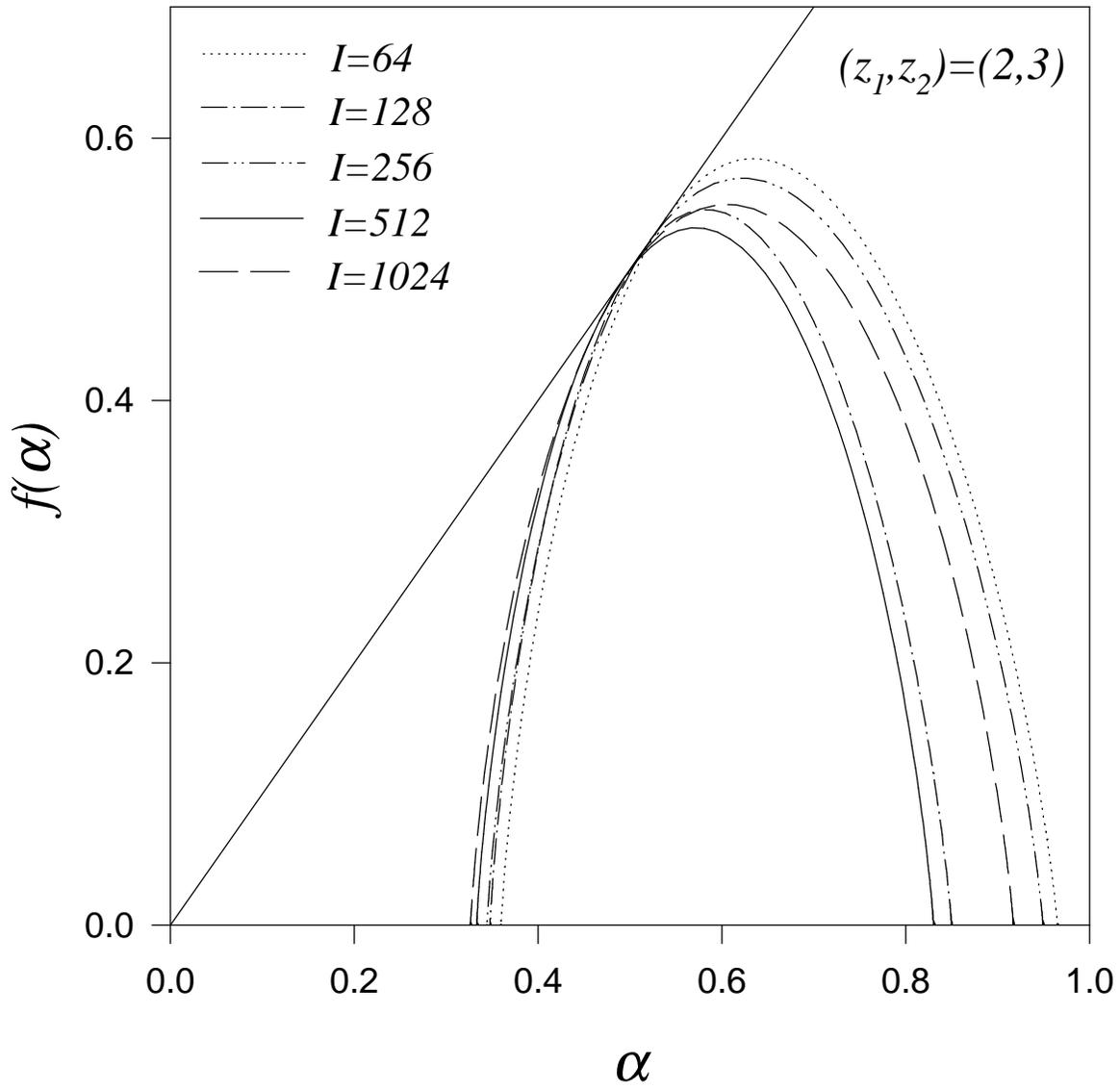
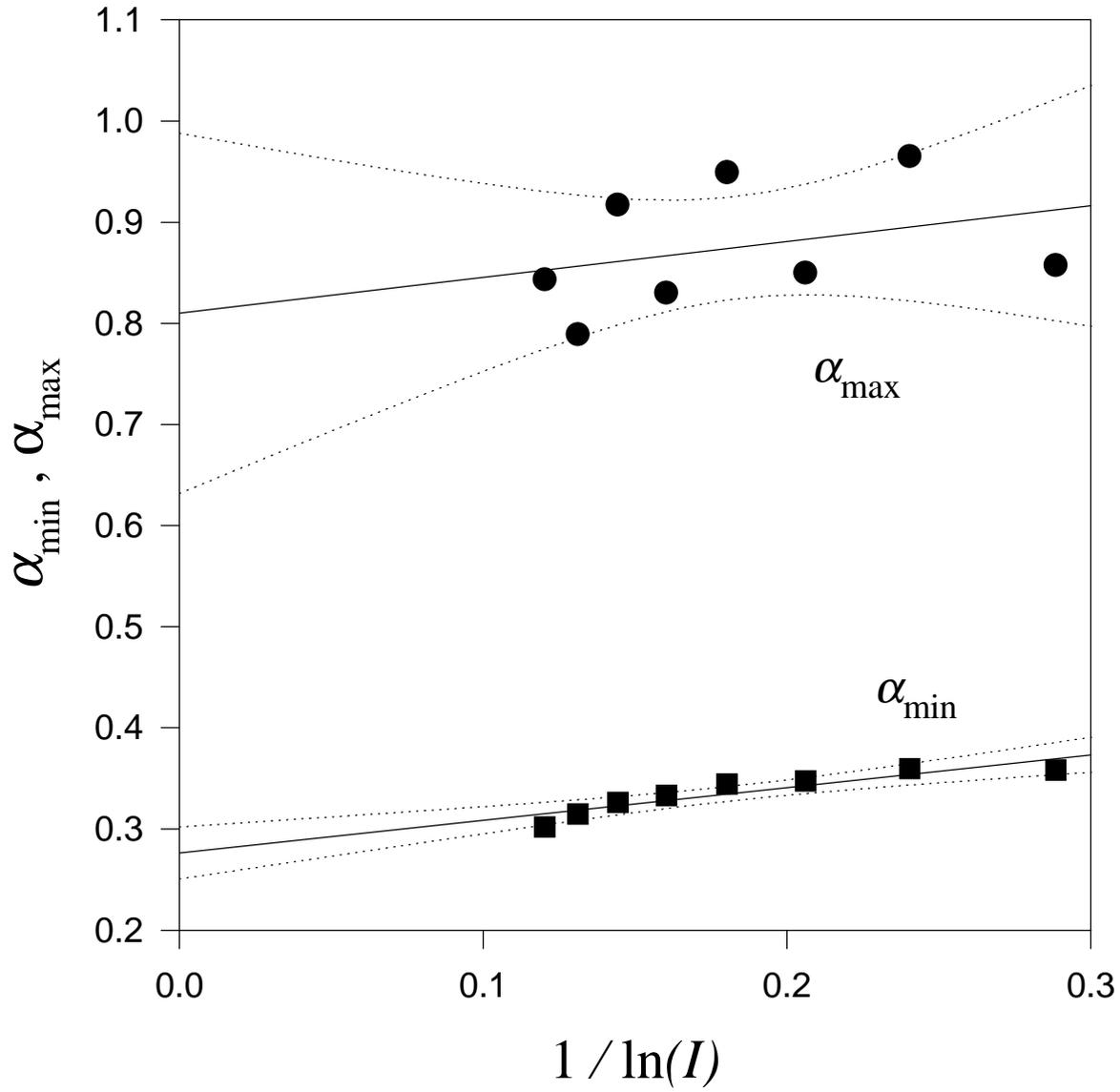
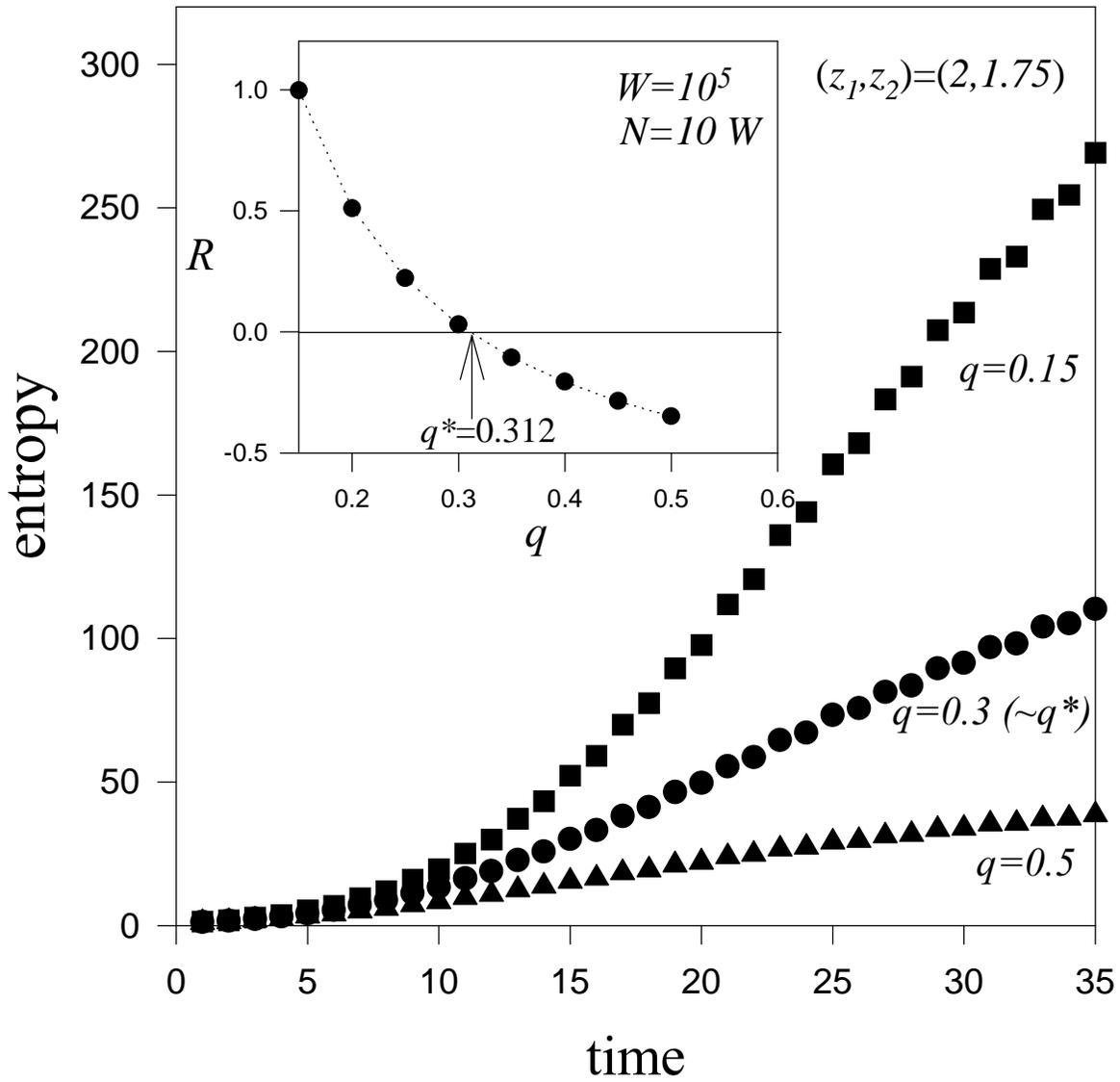
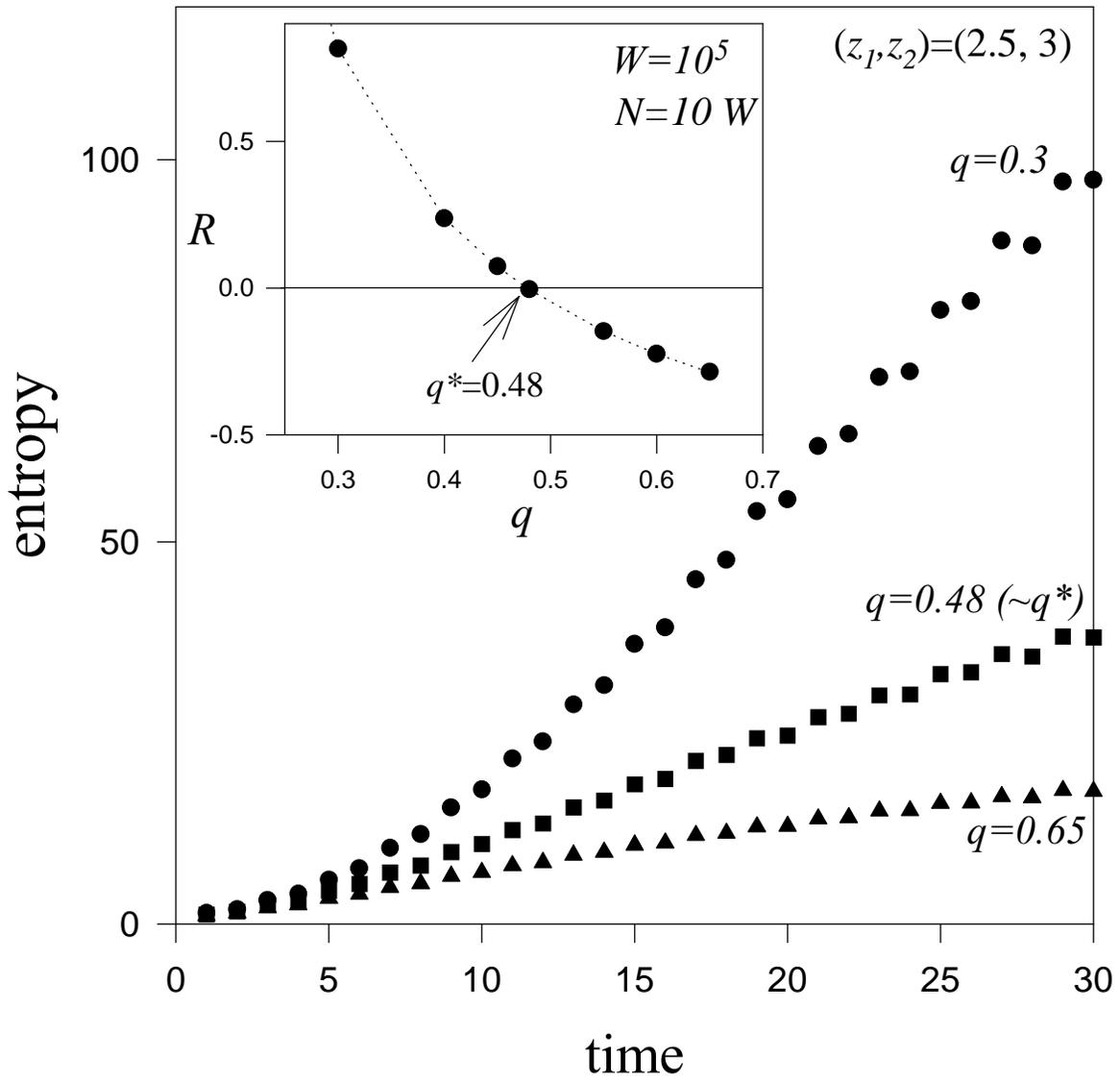
\end{document}